# An efficient certificateless authenticated key agreement protocol without bilinear pairings

Debiao He*, Yitao Chen

*School of Mathematics and Statistics, Wuhan University, Wuhan, China*

*Correspond author
Email: hedebiao@163.com
Tel: +008615307184927

**Abstract**: Certificateless public key cryptography simplifies the complex certificate management in the traditional public key cryptography and resolves the key escrow problem in identity-based cryptography. Many certificateless authenticated key agreement protocols using bilinear pairings have been proposed. But the relative computation cost of the pairing is approximately twenty times higher than that of the scalar multiplication over elliptic curve group. Recently, several certificateless authenticated key agreement protocols without pairings were proposed to improve the performance. In this paper, we propose a new certificateless authenticated key agreement protocol without pairing. The user in our just needs to compute five scale multiplication to finish the key agreement. We also show the proposed protocol is secure in the random oracle model.

**Classification Codes**: 11T71, 94A60

# 1. Introduction

Public key cryptography is an important technique to realize network and information security. Traditional public key infrastructure requires a trusted certification authority to issue a certificate binding the identity and the public key of an entity. Hence, the problem of certificate management arises. To solve the problem, Shamir defined a new public key paradigm called identity-based public key cryptography [1]. However, identity-based public key cryptography needs a trusted KGC to generate a private key for an entity according to his identity. So we are confronted with the key escrow problem. Fortunately, the two problems in traditional public key infrastructure and identity-based public key cryptography can be prohibited by introducing certificateless public key cryptography (CLPKC) [2], which can be conceived as an intermediate between traditional public key infrastructure and identity-based cryptography.

The first certificateless two-party authenticated key agreement(CTAKA) protocol appears in the seminal paper by Al-Riyami and Pa-terson [2]. However, no formal security model or proof for this CTAKA protocol is provided. Some early certificateless key exchange protocols (e.g., [3-6]) are proposed with heuristic security analysis. In order to improve the security, Swanson [7] proposed the first formal security model for the CTAKA protocol. Swanson also pointed that several early proposed CTAKA protocols[3-6] are insecure in his model. In [8], Lippold et al. proposed a new security model for CTAKA protocol. They also proposed a CTAKE protocol and prove its security under their model. Compared with the model by Swanson, Lippold et al.'s model is stronger in the sense that after the adversary replaces the public key of a user, the user will use the new public/private key pair in the rest of the game, while in Swanson's model, the user keeps using his/her original public/private key pair. However, the performance of Lippold et al.'s protocol is unacceptable. Very recently, Zhang et al.[9] proposed a different security model. They also proposed an efficient CTAKA protocol and demonstrated that their protocol is probably secure in their model.

All the above CTAKA protocols [2-9] are from bilinear pairings and the pairing is regarded as an expensive cryptography primitive. The relative computation cost of a pairing is approximately twenty times higher than that of the scalar multiplication over elliptic curve group [10]. Therefore, CTAKA protocols without bilinear pairings would be more appealing in terms of efficiency. Recently, several certificateless key exchange protocols without pairing have been proposed in [11-14]. However, Yang et al.[13] pointed both of Geng et al.'s protocol[11] and Hou et al.'s protocol[12] are not secure. They proposed an improved CTAKA protocol. He et al. [14] also proposed an CTAKA protocol without pairing. Unfortunately, Han [15] demonstrated that their scheme is not secure against the type 1 adversary.

In this paper, we propose a new CTAKA protocol without pairings. The user in our protocol just needs to compute five elliptic curve scale multiplications to end the key agreement. Then our protocol has the best performance among the CTAKA protocols. We also show our protocol is provably secure under the random oracle model.

The remainder of this paper is organized as follows. Section 2 describes some preliminaries. In Section 3, we propose our certificateless authenticated key agreement protocol. The security analysis of the proposed protocol is presented in Section 4. In Section 5, performance analysis is presented. Conclusions are given in Section 6.

## 2. Preliminaries

### 2.1 Background of elliptic curve group

Let the symbol $E/F_p$ denote an elliptic curve $E$ over a prime finite field $F_p$, defined by an equation

$$y^2 = x^3 + ax + b, \quad a,b \in F_p \tag{1}$$

and with the discriminant

$$\Delta = 4a^3 + 27b^2 \neq 0. \tag{2}$$

The points on $E/F_p$ together with an extra point $O$ called the point at infinity form a group

$$G = \{(x, y) : x, y \in F_p, E(x, y) = 0\} \cup \{O\}. \tag{3}$$

Let the order of $G$ be $n$. $G$ is a cyclic additive group under the point addition "+" defined as follows: Let $P, Q \in G$, $l$ be the line containing $P$ and $Q$ (tangent line to $E/F_p$ if $P = Q$), and $R$, the third point of intersection of $l$ with $E/F_p$. Let $l'$ be the line connecting $R$ and $O$. Then $P$ "+" $Q$ is the point such that $l'$ intersects $E/F_p$ at $R$ and $O$ and P "+" Q. Scalar multiplication over $E/F_p$ can be computed as follows:

$$tP = P + P + \cdots + P(t \ times) \qquad (4).$$

The following problems defined over $G$ are assumed to be intractable within polynomial time.

**Computational Diffie-Hellman (CDH) problem**: Given a generator $P$ of $G$ and $(aP, bP)$ for unknown $a, b \in_R Z_n^*$, compute $abP$. The CDH assumption states that the probability of any polynomial-time algorithm to solve the CDH problem is negligible.

## 2.2 CTAKA protocol

A CTAKA protocol consists of six polynomial-time algorithms[2, 8]: *Setup*, *Partial-Private-Key-Extract*, *Set-Secret-Value*, *Set-Private-Key*, *Set-Public-Key* and *Key-Agreement*. These algorithms are defined as follows.

*Setup*: **This algorithm takes** security parameter $k$ as input and returns the system parameters *params* and master key.

*Partial-Private-Key-Extract*: This algorithm takes *params*, master key and a user's identity $ID_i$ as inputs and returns a partial private key $D_i$.

*Set-Secret-Value*: This algorithm takes *params* and a user's identity $ID_i$ as inputs, and generates a secret value $x_i$.

*Set-Private-Key*: This algorithm takes *params*, a user's partial private key $D_i$ and his secret value $x_i$ as inputs, and outputs the full private key $S_i$.

*Set-Public-Key*: This algorithm takes *params* and a user's secret value $x_i$ as inputs, and generates a public key $P_i$ for the user.

*Key-Agreement*: This is a probabilistic polynomial-time interactive algorithm which involves two entities $A$ and $B$. The inputs are the system parameters *params* for both $A$ and $B$, plus $(S_A, ID_A, P_A)$ for $A$, and $(S_B, ID_B, P_B)$ for $B$. Here, $S_A$, $S_B$ are the respective private keys of $A$ and $B$; $ID_A$ is the identity of $A$ and $ID_B$ is the identity of $B$; $P_A$, $P_B$ are the respective public key of $A$ and $B$. Eventually, if the protocol does not fail, $A$ and $B$ will obtain a secret session key $K_{AB} = K_{BA} = K$.

## 2.3 Security model for CTAKA protocols

In CTAKA, as defined in [2], there are two types of adversaries with different capabilities, we assume *Type 1 Adversary*, $\mathcal{A}1$ acts as a dishonest user while *Type 2 Adversary*, $\mathcal{A}2$ acts as a malicious KGC:

**Type 1 Adversary**: Adversary $\mathcal{A}1$ does not have access to the master key, but $\mathcal{A}1$ can replace the public keys of any entity with a value of his choice, since there is no certificate involved in CLPKC.

**Type 2 Adversary**: Adversary $\mathcal{A}2$ has access to the master key, but cannot replace any user's public key.

Very recently, Zhang et al.'s [8] present a security model for AKA protocols in the setting of CLPKC. The model is defined by the following game between a challenger $\mathcal{C}$ and an adversary $\mathcal{A} \in \{\mathcal{A}1, \mathcal{A}2\}$. In their et al.'s model, $\mathcal{A}$ is modeled by a probabilistic polynomial-time turing machine. All communications go through the adversary $\mathcal{A}$. Participants only respond to the queries by $\mathcal{A}$ and do not communicate directly among themselves. $\mathcal{A}$ can relay, modify, delay, interleave or delete all the message flows in the system. Note that $\mathcal{A}$ can act as a benign adversary, which means that $\mathcal{A}$ is deterministic and restricts her action to choosing a pair of oracles $\prod_{i,j}^{n}$ and $\prod_{j,i}^{t}$ and then faithfully conveying each message flow from one oracle to the other. Furthermore, $\mathcal{A}$ may ask a polynomially bounded number of the following queries as follows.

*Create*($ID_i$): This allows $\mathcal{A}$ to ask $\mathcal{C}$ to set up a new participant i with identity $ID_i$. On receiving such a query, $\mathcal{C}$ generates the public/private key pair for $i$.

*Public – Key*($ID_i$): $\mathcal{A}$ can request the public key of a participant $i$ whose identity is $ID_i$. To respond, $\mathcal{C}$ outputs the public key $P_i$ of participant $i$.

*Partial - Private - Key*($ID_i$): $\mathcal{A}$ can request the partial private key of a participant $i$ whose identity is $ID_i$. To respond, $\mathcal{C}$ outputs the partial private key $D_i$ of participant $i$.

*Corrupt*($ID_i$): $\mathcal{A}$ can request the private key of a participant $i$ whose identity is $ID_i$. To respond, $\mathcal{C}$ outputs the private key $S_i$ of participant $i$.

*Public − Key − Replacement*$(ID_i, P'_i)$: For a participant $i$ whose identity is $ID_i$; $\mathcal{A}$ can choose a new public key $P'$ and then set $P'$ as the new public key of this participant. $\mathcal{C}$ will record these replacements which will be used later.

*Send*$(\prod_{i,j}^{n}, M)$: $\mathcal{A}$ can send a message $M$ of her choice to an oracle, say $\prod_{i,j}^{n}$, in which case participant $i$ assumes that the message has been sent by participant $j$. $\mathcal{A}$ may also make a special *Send* query with $M \neq \lambda$ to an oracle $\prod_{i,j}^{n}$, which instructs $i$ to initiate a protocol run with $j$. An oracle is an initiator oracle if the first message it has received is $\lambda$. If an oracle does not receive a message $\lambda$ as its first message, then it is a responder oracle.

*Reveal*$(\prod_{i,j}^{n})$: $\mathcal{A}$ can ask a particular oracle to reveal the session key (if any) it currently holds to $\mathcal{A}$.

*Test*$(\prod_{i,j}^{n})$: At some point, $\mathcal{A}$ may choose one of the oracles, say $\prod_{I,J}^{T}$, to ask a single *Test* query. This oracle must be fresh. To answer the query, the oracle flips a fair coin $b \in \{0,1\}$, and returns the session key held by $\prod_{I,J}^{T}$ if $b = 0$, or a random sample from the distribution of the session key if $b = 1$.

After a *Test* query, the adversary can continue to query the oracles except that it cannot make a Reveal query to the test oracle $\prod_{I,J}^{T}$ or to $\prod_{J,I}^{t}$ who has a matching conversation with $\prod_{I,J}^{T}$ (if it exists), and it cannot corrupt participant $J$. In addition, if $\mathcal{A}$ is a Type 1 adversary, $\mathcal{A}$ cannot request the partial private key of the participant $J$; and if $\mathcal{A}$ is a Type 2 adversary, $J$ cannot replace the public key of the participant $J$. At the end of the game, $\mathcal{A}$ must output a guess bit $b'$. $\mathcal{A}$ wins if and only if $b' = b$. $\mathcal{A}$'s advantage to win the above game, denoted by $Advantage^{A}(k)$, is defined as: $Advantage^{A}(k) = \left| \Pr[b' - b] - \frac{1}{2} \right|$.

**Definition 1**. A CTAKA protocol is said to be secure if:

(1) In the presence of a benign adversary on $\prod_{i,j}^{n}$ and $\prod_{j,i}^{t}$, both oracles always agree on the same session key, and this key is distributed uniformly at random.

(2) For any adversary, $Advantage^{A}(k)$ is negligible.

## 3. Our protocol

In this section, we will propose a new CTAKA protocol. Our protocol consists of six polynomial-time algorithms. They are described as follows.

**Setup:** This algorithm takes a security parameter $k$ as in put, and returns system parameters and a master key. Given $k$, KGC does the following.

1) KGC chooses a $k$-bit prime $p$ and determines the tuple $\{F_p, E/F_p, G, P\}$ as defined in Secttion 2.1.
2) KGC chooses the master private key $s \in Z_n^*$ and computes the master public key $P_{pub} = sP$.
3) KGC chooses two cryptographic secure hash functions $H_1 : \{0,1\}^* \to Z_n^*$ and $H_2 : \{0,1\}^* \to Z_n^*$.
4) KGC publishes $params = \{F_p, E/F_p, G, P, P_{pub}, H_1, H_2\}$ as system parameters and secretly keeps the master key $s$.

**Set-Secret-Value**: The user with identity $ID_i$ picks randomly $x_i \in Z_n^*$, computes $P_i = x_i \cdot P$ and sets $x_i$ as his secret value.

**Partial-Private-Key-Extract:** This algorithm takes master key, a user's identifier, $P_i$, system parameters as input, and returns the user's ID-based private key. With this algorithm, for each user with identifier $ID_i$, KGC works as follows.

1) KGC chooses a random number $r_i \in Z_n^*$, computes $R_i = r_i \cdot P$ and $h_i = H_1(ID_i, R_i, P_i)$.
2) KGC computes $s_i = r_i + h_i s \bmod n$ and issues $\{s_i, R_i\}$ to the users through secret channel.

The user's s partial private key is the tuple $s_i$ and he can validate her private key by checking whether the equation $s_i \cdot P = R_i + h_i \cdot P_{pub}$ holds. The private key is valid if the equation holds and vice versa.

**Set-Private-Key**: The user with identity $ID_i$ takes the pair $sk_i = (x_i, s_i)$ as its private key.

**Set-Public-Key**: The user with identity $ID_i$ takes $pk_i = \{P_i, R_i\}$ as its public key.

**Key-Agreement**: Assume that an entity $A$ with identity $ID_A$ has private key $sk_A = (x_A, s_A)$ and public key $pk_A = \{P_A, R_A\}$ and an entity $B$ with identity $ID_B$ has private key $sk_B = (x_B, s_B)$ and public key $pk_B = \{P_B, R_B\}$ want to establish a session key, then they can do, as shown in Fig.1, as follows.

1) $A$ chooses a random number $a \in Z_n^*$ and computes $T_A = a \cdot P$, then $A$ send $M_1 = \{ID_A, T_A\}$ to $B$.

2) After receiving $M_1$, $B$ chooses a random number $b \in Z_n^*$ and computes $T_B = b \cdot P$, then $B$ send $M_2 = \{ID_B, T_B\}$ to $A$.

Then both $A$ and $B$ can compute the shared secrets as follows.

$A$ computes

$$K_{AB}^1 = (x_A + s_A)T_B + a \cdot (P_B + R_B + H_1(ID_B, R_B, P_B)P_{pub}) \text{ and } K_{AB}^2 = a \cdot T_B \quad (5)$$

$B$ computes

$$K_{BA}^1 = (x_B + s_B)T_A + b \cdot (P_A + R_A + H_1(ID_A, R_A, P_A)P_{pub}) \text{ and } K_{BA}^2 = b \cdot T_A \quad (6)$$

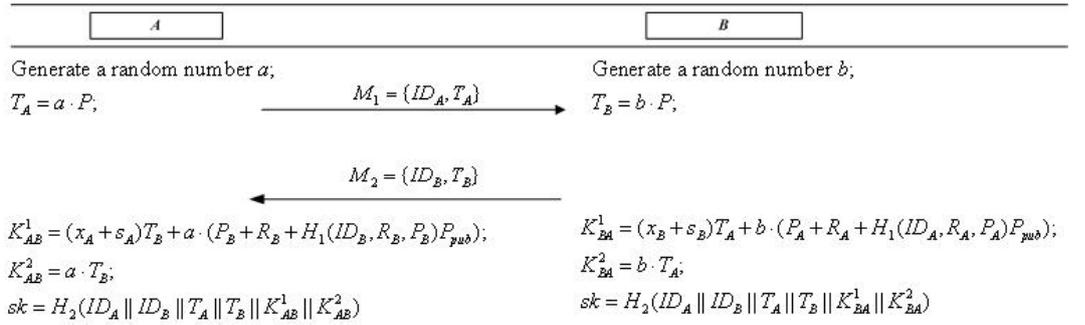

Fig. 1. Key agreement of our protocol

The shared secrets agree because:

$$\begin{aligned} K_{AB}^1 &= (x_A + s_A)T_B + a \cdot (P_B + R_B + H_1(ID_B, R_B, P_B)P_{pub}) \\ &= (x_A + s_A)T_B + a(x_B + s_B)P = (x_A + s_A)T_B + (x_B + s_B)T_A \\ &= b \cdot (P_A + R_A + H_1(ID_A, R_A, P_A)P_{pub})P + (x_B + s_B)T_A \\ &= K_{BA}^1 \end{aligned} \quad (7)$$

and

$$K_{AB}^2 = abP = baP = K_{BA}^2 \quad (8)$$

Thus the agreed session key for $A$ and $B$ can be computed as:

$$\begin{aligned} sk &= H_2(ID_A \| ID_B \| T_A \| T_B \| K_{AB}^1 \| K_{AB}^2) \\ &= H_2(ID_A \| ID_B \| T_A \| T_B \| K_{BA}^1 \| K_{BA}^2) \end{aligned} \quad (9)$$

## 4. Security Analysis

To prove the security of our protocol in the random oracle model, we treats $H_1$ and $H_2$ as two random oracles [16] using the model defined in [9]. For the security, the following lemmas and theorems are provided.

**Lemma 1**. If two oracles are matching, both of them will be accepted and will get the same session key which is distributed uniformly at random in the session key sample space.

*Proof.* From the correction analysis of our protocol in section 4.1, we know if two oracles are matching, then both of them are accepted and have the same session key. The session key is distributed uniformly since $a$ and $b$ are selected uniformly during the protocol execution.

**Lemma 2**. Assuming that the CDH problem is intractable, the advantage of a Type 1 adversary against our protocol is negligible in the random oracle model.

*Proof.* Suppose that there is a Type 1 Adversary $\mathcal{A}1$ who can win the game defined in Section 2 with a non-negligible advantage $Advantage^A(k)$ in polynomial-time $t$. Then, $\mathcal{A}1$ can win the game with non-negligible probability $\varepsilon$, we show how to use the ability of $\mathcal{A}1$ to construct an algorithm $\mathcal{C}$ to solve the CDH problem.

Suppose $\mathcal{C}$ is given an instance $(aP, bP)$ of the CDH problem, and wants to compute $cP$ with $c = ab \bmod n$. $\mathcal{C}$ first chooses $P_0 \in G$ at random, sets $P_0$ as the system public key $P_{pub}$, selects the system parameter $params = \{F_p, E/F_p, G, P, P_{pub}, H_1, H_2\}$, and sends $params$ to $\mathcal{A}1$. Let $q_s$ be the maximal number of sessions each participant may be involved in. Supposed $\mathcal{A}1$ makes at most $q_{H_i}$ times $H_i$ queries and creates at most $q_c$ participants. $\mathcal{C}$ chooses at random $I, J \in [1, q_{H_1}]$, $T \in [1, q_s]$, and answers $\mathcal{A}1$'s queries as follows.

$Create(ID_i)$: $\mathcal{C}$ maintains an initially empty list $L_C$ consisting of tuples of the form ($ID_i, D_i, x_i, P_i$). If $ID_i = ID_I$, $\mathcal{C}$ chooses a random $x_i, h_i \in Z_n^*$ and computes $R_i = bP - h_i P_0$, public key $P_i = x_i P$, then $i$'s partial private key, private key and public key are $\perp$, $sk_i = (x_i, \perp)$ and $pk_i = (P_i, R_i)$ separately.

Otherwise, $\mathcal{C}$ chooses a random $x_i, s_i, h_i \in Z_n^*$ and computes $R_i = s_i P - h_i P_0$, $P_i = x_i P$, then $i$'s partial private key, private key and public key are $s_i$, $sk_i = (x_i, s_i)$ and $pk_i = (P_i, R_i)$ separately. At last, $\mathcal{C}$ adds the tuple $(ID_i, R_i, P_i, h_i)$ and $(ID_i, s_i, sk_i, pk_i)$ to the list $L_{H_1}$ and $L_C$, separately.

$H_1(ID_i, R_i, P_i)$: $\mathcal{C}$ maintains an initially empty list $L_{H_1}$ which contains tuples of the form $(ID_i, R_i, P_i, h_i)$. If $(ID_i, R_i, P_i)$ is on the list $L_{H_1}$, then returns $h_i$. Otherwise, $\mathcal{C}$ executes the query $Create(ID_i)$ and returns $h_i$.

$Public-Key(ID_i)$: On receiving this query, $\mathcal{C}$ first searches for a tuple $(ID_i, s_i, sk_i, pk_i)$ in $L_C$ which is indexed by $ID_i$, then returns $pk_i$ as the answer.

$Partial-Private-Key(ID_i)$: Whenever $\mathcal{C}$ receives this query, if $ID_i = ID_I$ $\mathcal{C}$ aborts; else, $\mathcal{C}$ searches for a tuple $(ID_i, s_i, sk_i, pk_i)$ in $L_C$ which is indexed by $ID_i$ and returns $sk_i$ as the answer.

$Corrupt(ID_i)$: Whenever $\mathcal{C}$ receives this query, if $ID_i = ID_I$ $\mathcal{C}$ aborts. Otherwise, $\mathcal{C}$ searches for a tuple $(ID_i, s_i, sk_i, pk_i)$ in $L_C$ which is indexed by $ID_i$ and if $x_i = null$, $\mathcal{C}$ returns *null*. Otherwise, $\mathcal{C}$ returns $(s_i, sk_i)$ as the answer.

$Public-Key-Replacement(ID_i, pk_i')$: On receiving this query, $\mathcal{C}$ searches for a tuple $(ID_i, s_i, sk_i, pk_i)$ in $L_C$ which is indexed by $ID_i$, then updates $pk_i$ to $pk_i'$ and sets $s_i = \perp, sk_i = \perp$.

$Send(\prod_{i,j}^n, M)$: $\mathcal{C}$ maintains an initially empty list $L_S$ consisting of tuples of the form $(\prod_{i,j}^n, trans_{i,j}^n, r_{i,j}^n)$, where $trans_{i,j}^n$ is the transcript of $\prod_{i,j}^n$ so far and $r_{i,j}^n$ will be described later. $\mathcal{C}$ answers the query as follows:

- If $n = T$, $ID_i = ID_I$ and $ID_j = ID_J$, $\mathcal{C}$ returns $aP$ as the answer and updates the tuple $(\prod_{i,j}^n, trans_{i,j}^n, r_{i,j}^n)$ $r_{i,j}^n = \perp$.
- Otherwise, $\mathcal{C}$ answers the query according to the specification of the protocol. Note that when $M$ is not the second message to $\prod_{i,j}^n$, $\mathcal{C}$ chooses at random $r_{i,j}^n \in Z_n^*$ and computes $r_{i,j}^n P$ as the reply. Then $\mathcal{C}$ updates the tuple indexed by $\prod_{i,j}^n$ in $L_S$.

$Reveal(\prod_{i,j}^n)$ : $\mathcal{C}$ maintains a list $L_R$ of the form $(\prod_{i,j}^n, ID_{ini}^n, ID_{resp}^n, T_{ini}^n, T_{resp}^n, SK_{i,j}^n)$ where $ID_{ini}^n$ is the identification of the initiator in the session which $\prod_{i,j}^n$ engages in and $ID_{resp}^n$ is the identification of the responder. $\mathcal{C}$ answers the query as follows:

- If $n = T$, $ID_i = ID_I$ and $ID_j = ID_J$ or $\prod_{i,j}^n$ is the oracle who has a matching conversion with $\prod_{I,J}^T$, $\mathcal{C}$ aborts.
- Else if $ID_i \neq ID_I$,
  - $\mathcal{C}$ looks up the list $L_S$ and $L_C$ for corresponding tuple $(\prod_{i,j}^n, r_{i,j}^n, T_{i,j}^n, T_{j,i}^n, R_i^n, R_j^n, P_i^n, P_j^n)$ and $(ID_i, D_i, x_i, P_i)$ separately. Then $\mathcal{C}$ computes $K_{i,j}^1 = (x_i + s_i)T_{j,i}^n + r_{j,i}^n(P_j^n + R_j^n + H_1(ID_j \| R_j^n)P_{pub})$, $K_{i,j}^1 = r_{j,i}^n T_{j,i}^n$.
  - $\mathcal{C}$ makes a $H_2$ query. If $\prod_{i,j}^n$ is the initiator oracle then the query is of the form $(ID_i \| ID_j \| T_i \| T_j \| K_{i,j}^1 \| K_{i,j}^2)$ or else of the form $(ID_j \| ID_i \| T_j \| T_i \| K_{i,j}^1 \| K_{i,j}^2)$.
- Else ($ID_i = ID_I$),
  - $\mathcal{C}$ looks up the list $L_S$ for corresponding tuple $(\prod_{i,j}^n, r_{i,j}^n, T_{i,j}^n, T_{j,i}^n, R_i^n, R_j^n, P_i^n, P_j^n)$.
  - $\mathcal{C}$ looks up the list $L_{H_2}$ to see if there exists a tuple index by $(ID_i, ID_j, T_i, T_j)$. If $\prod_{i,j}^n$ is an initiator, otherwise index by $(ID_j, ID_i, T_j, T_i)$.
  - If there exists such tuple and the corresponding $K_{i,j}^1$ and $K_{i,j}^2$ satisfies the equation $e(K_{i,j}^2, P) = e(T_i^n, T_j^n)$ and $e(K_{i,j}^1 - r_{i,j}^n(P_j^n + R_j^n + H_1(ID_j \| R_j^n \| P_j^n)P_{pub}, P) = e(P_i^n + R_i^n + H_1(ID_i \| R_i^n \| P_i^n)P_{pub}, T_j^n)$ given a proper bilinear map $e$ for group $G$, then $\mathcal{C}$ obtains the corresponding $h_i$ and sets $SK_{i,j}^n = h_i$. Otherwise $\mathcal{C}$ chooses at random $SK_{i,j}^n \in \{0,1\}^k$.

$H_2$ query: $\mathcal{C}$ maintains a list $L_{H_2}$ of the form $(ID_u^i, ID_u^j, T_u^i, T_u^j, K_u^1, K_u^2, h_u)$ and A responds with $H_2$ queries $(ID_u^i, ID_u^j, T_u^i, T_u^j, K_u^1, K_u^2)$ as follows:

- If a tuple indexed by $(ID_u^i, ID_u^j, T_u^i, T_u^j, K_u^1, K_u^2)$ is already in $L_{H_2}$, $\mathcal{C}$ replies with the corresponding $h_u$.
- Else, if there is no such a tuple,
  - If the equation $e(K_u^2, P) = e(T_u^i, T_u^j)$ and $e(K_u^1, P) = e(P_i + R_i + H_1(ID_i \| R_i \| P_i)P_{pub}, T_u^j)e(P_j + R_j + H_1(ID_j \| R_j \| P_j)P_{pub}, T_u^i)$ hold given a proper bilinear pairing e for group $G$, go through the list $L_R$. If there is such a tuple indexed by $(ID_u^i, ID_u^j, T_u^i, T_u^j)$ in the

list $L_R$, then $\mathcal{C}$ obtains the corresponding $SK_{i,j}^n$ and sets $SK_{i,j}^n = h_u$. Otherwise $\mathcal{C}$ chooses at random $h_u \in \{0,1\}^k$.

✧ Else if the equations do not hold for $(ID_u^i, ID_u^j, T_u^i, T_u^j, K_u^1, K_u^2)$, $\mathcal{C}$ chooses at random $h_u \in \{0,1\}^k$.

✧ $\mathcal{C}$ inserts the tuple $(ID_u^i, ID_u^j, T_u^i, T_u^j, K_u^1, K_u^2, h_u)$ into the list $L_{H_2}$.

$Test(\prod_{I,J}^T)$: At some point, $\mathcal{C}$ will ask a *Test* query on some oracle. If $\mathcal{C}$ does not choose one of the oracles $\prod_{I,J}^T$ to ask the *Test* query, then $\mathcal{C}$ aborts. Otherwise, $\mathcal{C}$ simply outputs a random value $x \in \{0,1\}^k$.

The probability that $\mathcal{C}$ chooses $\prod_{I,J}^T$ as the *Test* oracle and that $\frac{1}{q_C^2 q_s}$. In this case, $\mathcal{C}$ would not have made $Corrupt(\prod_{I,J}^T)$ or $Reveal(\prod_{I,J}^T)$ queries, and so $\mathcal{C}$ would not have aborted. If $\mathcal{C}$ can win in such a game, then $\mathcal{C}$ must have made the corresponding $H2$ query of the form $(ID_T^i, ID_T^j, T_T^i, T_T^j, K_T^1, K_T^2)$. If $\prod_{I,J}^T$ is the initiator oracle or else $(ID_T^j, ID_T^i, T_T^j, T_T^i, K_T^1, K_T^2)$ with overwhelming probability because $H_2$ is a random oracle. Thus $\mathcal{C}$ can find the corresponding item in the $H_2$-list with the probability $\frac{1}{q_{H_2}}$ and output $K_T^1 - (x_I - h_I)(aP) - r_{I,J}^T(P_J + R_J + h_J P_{pub})$ as a solution to the CDH problem. The probability that $\mathcal{C}$ solves the CDH problem is $\frac{\varepsilon}{q_C^2 q_s q_{H_2}}$.

**Lemma 3**. Under the assumption that the CDH problem is intractable, the advantage of a Type 2 adversary against our protocol is negligible in the random oracle model.

*Proof*. Suppose that there is a Type 2 adversary $\mathcal{A}2$ who can win the game defined in Section 2 with a non-negligible advantage $Advantage^A(k)$ in polynomial-time $t$. Then, $\mathcal{A}2$ can win the game with no-negligible probability $\varepsilon$, we show how to use the ability of $\mathcal{A}2$ to construct an algorithm $\mathcal{C}$ to solve the CDH problem.

Suppose $\mathcal{C}$ is given an instance $(aP, bP)$ of the CDH problem, and want to compute $cP$ with $c = ab \bmod n$. $\mathcal{C}$ first chooses $sP \in G$ at random, sets $sP$ as the system public key $P_{pub}$, selects the system parameter $params = \{F_p, E/F_p, G, P, P_{pub}, H_1, H_2\}$, and sends $params$ and master key $s$ to $\mathcal{A}2$. Let $q_s$ be the maximal number of sessions each participant may be involved in. Supposed $\mathcal{A}2$ makes at most $q_{H_i}$ times $H_i$ queries and creates at most $q_c$ participants. $\mathcal{C}$ chooses at random $I, J \in [1, q_{H_1}]$, $T \in [1, q_s]$, and answers $\mathcal{A}2$'s queries as follows.

$Create(ID_i)$: $\mathcal{C}$ maintains an initially empty list $L_C$ consisting of tuples of the form ($ID_i, s_i, sk_i, pk_i$). If $ID_i = ID_I$, $\mathcal{C}$ chooses a random $r_i, h_i \in Z_n^*$ and computes $R_i = r_i P$, $s_i = r_i + h_i s \mod n$, $P_i = bP$, then $i$'s partial private key, private key and public key are $s_i$, $sk_i = (\bot, s_i)$ and $pk_i = \{P_i, R_i\}$ separately. Otherwise, $\mathcal{C}$ chooses a random $x_i, r_i, h_i \in Z_n^*$ and computes $R_i = r_i P$, $s_i = r_i + h_i s \mod n$, public key $P_i = x_i P$, then $i$'s partial private key, private key and public key are $s_i$, $sk_i = (x_i, s_i)$ and $pk_i = \{P_i, R_i\}$ separately. At last, $\mathcal{C}$ add the tuple ($ID_i, R_i, P_i, h_i$) and ($ID_i, s_i, sk_i, pk_i$) to the list $L_{H_1}$ and $L_C$, separately.

$\mathcal{C}$ answers $\mathcal{A}2$'s $H_1(ID_i, R_i, P_i)$, $Public-Key(ID_i)$, $Corrupt(ID_i)$, $Partial-Private-Key(ID_i)$, $Send(\prod_{i,j}^n, M)$, $Reveal(\prod_{i,j}^n)$, $H_2$ query and $Test(\prod_{I,J}^T)$ queries like he does in lemma 2.

The probability that $\mathcal{C}$ chooses $\prod_{I,J}^T$ as the $Test$ oracle and that $\frac{1}{q_C^2 q_s}$. In this case, $\mathcal{C}$ would not have made $Corrupt(\prod_{I,J}^T)$ or $Reveal(\prod_{I,J}^T)$ queries, and so $\mathcal{C}$ would not have aborted. If $\mathcal{C}$ can win in such a game, then $\mathcal{C}$ must have made the corresponding H2 query of the form ($ID_T^i, ID_T^j, T_T^i, T_T^j, K_T^1, K_T^2$) if $\prod_{I,J}^T$ is the initiator oracle or else ($ID_T^j, ID_T^i, T_T^j, T_T^i, K_T^1, K_T^2$) with overwhelming probability because $H_2$ is a random oracle. Thus $\mathcal{C}$ can find the corresponding item in the $H_2$-list with the probability $\frac{1}{q_{H_2}}$ and output $K_T^1 - s_I(bP) - r_{I,J}^T(P_J + R_J + h_J P_{pub})$ as a solution to the CDH problem. The probability that $\mathcal{C}$ solves the CDH problem is $\frac{\varepsilon}{q_C^2 q_s q_{H_2}}$.

From the above three lemmas, we can get the following two theorems.

**Theorem 1**. Our protocol is a secure CTAKA protocol.

Through the similar method, we can prove our protocol could provide forward secrecy property. We will describe it in the following theorem.

**Theorem 2**. Our protocol has the perfect forward secrecy property if the CDH problem in $G$ is hard.

## 5. Comparison with previous protocol

For the convenience of evaluating the computational cost, we define some notations as follows.

$T_{mul}$: The time of executing a scalar multiplication operation of point.

$T_{add}$: The time of executing an addition operation of points.

$T_{inv}$: The time of executing a modular invasion operation.

$T_h$: The time of executing a one-way hash function.

We will compare the efficiency of our new protocol with there CTAKA protocols without pairings, i.e. Geng et al.'s protocol [11], Hou et al.'s protocol [12], Yang et al.'s protocol[13], and He et al.'s protocol[14]. In Table 1, we summarize the performance results of the proposed user authentication and key exchange protocol.

**Table 1**. Comparison of different protocols

|  | Geng et al.'s protocol [11] | Hou et al.'s protocol [12] | Yang et al.'s protocol [13] | He et al's protocol[14] | Our protocol |
|---|---|---|---|---|---|
| Cost | $7T_{mul} + 2T_h$ | $6T_{mul} + 2T_h$ | $9T_{mul} + 2T_h$ | $5T_{mul} + 3T_{add}$ $+T_{inv} + 2T_h$ | $5T_{mul} + 4T_{add}$ $+2T_h$ |

As the main computational overheads, we only consider the scale multiplication. Then we can conclude the computational cost of our protocol is 71.43% of Geng et al.'s scheme [11], 83.33% of Hou et al.'s scheme[12], and 55.56% of Yang et al.'s scheme[13]. Moreover, Geng et al.'s protocol [11] and Hou et al.'s protocol[12] are not secure[13]. He et al.'s protocol [14] has almost the same performance as our protocol. But He et al.'s protocol [14] is not secure either [15]. Thus our scheme is more useful and efficient than the previous schemes.

# 6. Conclusion

The certificateless public key cryptography is receiving significant attention because it is a new paradigm that simplifies the public key cryptography. We then proposed a new CTAKA protocol without pairings and proved its security in the random oracle model under the CDH assumption. The proposed protocol has the best performance among the related protocols.

Many researchers have expressed doubts about the wisdom of relying on the random oracle model. In particular, Canetti et al. [17] proved that there are signature and encryption schemes which are secure in the random oracle model, but insecure for any instantiation of the standard oracle. To get better security, it is necessary to construct CTAKA protocol without pairings in the standard model. In the future, we will investigate the extraction algorithm for the standard model first. Then we will use the extraction algorithm to construct the CTAKA protocol without pairings in standard model such that it can be applied to more applications.

# References

[1]. A. Shamir, Identity-based cryptosystems and signature protocols, Proc. CRYPTO1984, LNCS, vol.196, 1984, pp.47–53.

[2]. S. Al-Riyami, K.G. Paterson, Certificateless public key cryptography, Proceedings of ASIACRYPT 2003, LNCS 2894, Springer-Verlag, 2003, pp. 452–473.

[3]. Z. Shao. Efficient authenticated key agreement protocol using self-certifed public keys from pairings. Wuhan University Journal of Natural Sciences, 10(1):267-270, 2005.

[4]. S. Wang, Z. Cao, X. Dong, Certificateless authenticated key agreement based on the MTI/CO protocol, Journal of Information and Computational Science 3 (2006) 575–581.

[5]. T. Mandt, C. Tan, Certificateless authenticated two-party key agreement protocols, in: Proceedings of the ASIAN 2006, LNCS, vol. 4435, Springer-Verlag, 2008, pp. 37–44.

[6]. Y. Shi, J. Li, Two-party authenticated key agreement in certificateless public key cryptography, Wuhan University Journal of Natural Sciences 12 (1) (2007) 71–74.

[7]. C. Swanson. Security in key agreement: Two-party certi_cateless schemes. Master Thesis, University of Waterloo, 2008.

[8]. G. Lippold, C. Boyd, J. Nieto. Strongly secure certificateless key agreement. In Pairing 2009, pages 206-230.

[9]. L. Zhang, F. Zhang, Q. Wua, J. Domingo-Ferrer, Simulatable certificateless two-party authenticated key agreement protocol, Information Sciences 180 (2010) 1020–1030.

[10]. L. Chen, Z. Cheng, and N.P. Smart, Identity-based key agreement protocols from pairings, Int. J. Inf. Secur., 6(2007) pp.213–241.

[11]. M. Geng and F. Zhang. Provably secure certificateless two-party authenticated key agreement protocol without pairing. In International Conference on Computational Intelligence and Security, pages 208-212, 2009.

[12]. M. Hou and Q. Xu. A two-party certificateless authenticated key agreement protocol without pairing. In 2nd IEEE International Conference on Computer Science and Information Technology, pages 412-416, 2009.

[13]. G. Yang, C. Tan, 6th ACM Symposium on Information, Computer and Communications Security, 71-79, 2011.

[14]. D. He, J. Chen, J. Hu, A pairing-free certificateless authenticated key agreement protocol, International Journal of Communication Systems, DOI: 10.1002/dac.1265, 2011.

[15]. W. Han, Breaking a certificateless key agreement protocol without bilinear pairing, http://eprint.iacr.org/2011/249.pdf

[16]. M. Bellare and P. Rogaway, Random oracles are practical: A paradigm for designing efficient protocols, in Proc. 1st ACM Conf. Comput. Commun. Security, 1993, pp. 62–73.

[17]. R. Canetti, O. Goldreich, S. Halevi. The random oracle methodology, revisited. Journal of ACM 2004; 51(4):557-594.